# Spin-dependent gain and loss in photonic quantum spin Hall systems

Tian-Rui Liu[1], Kai Bai[1], Jia-Zheng Li[1], Liang Fang[1], Duanduan Wan[1†] and Meng Xiao[1,2†]

[1] *Key Laboratory of Artificial Micro- and Nano-structures of Ministry of Education and School of Physics and Technology, Wuhan University, Wuhan 430072, China*

[2] *Wuhan Institute of Quantum Technology, Wuhan 430206, China*

Corresponding E-mail: † ddwan@whu.edu.cn; * phmxiao@whu.edu.cn

**Abstract**: Topological phases are greatly enriched by including non-Hermiticity. While most works focus on the topology of the eigenvalues and eigenstates, how topologically nontrivial non-Hermitian systems behave in dynamics has only drawn limited attention. Here, we consider a breathing honeycomb lattice known to emulate the quantum spin Hall effect and exhibits higher-order corner modes. We find that non-reciprocal intracell couplings introduce gain in one pseudo-spin subspace while loss with the same magnitude in the other. In addition, non-reciprocal intracell couplings can also suppress the spin mixture of the edge modes at the boundaries and delocalize the higher-order corner mode. Our findings deepen the understanding of non-Hermitian topological phases and bring in the spin degree of freedom in manipulating the dynamics in non-Hermitian systems.

# Introduction

Non-Hermitian systems [1–6] exhibit a wide range of counterintuitive phenomena that have widespread and profound applications in a variety of fields, such as photonics [7–11], acoustics [12–14], circuits [15,16], etc. [17,18] Non-Hermitian periodic systems have two types of bandgaps, i.e., point gaps and line gaps [19] and have more symmetries than Hermitian systems [20,21]. Thus, in principle, they should exhibit much more complex topological structures. The topological invariants of non-Hermitian systems have been defined with their eigenvalues [22,23] and eigenstates [14,23–27]. Currently, the vast majority of works focus on the topology of static systems. In contrast, the consequence induced by the nontrivial topology of periodic non-Hermitian systems in dynamics has only drawn limited attention [27,28]. Such dynamics are complicated since the eigenvalues are generally complex; thus, the contributions of different eigenstates can vary significantly during their evolution. On the other hand, the dynamics of non-Hermitian systems consisting of a few resonators have already been known to lead to novel phenomena such as mode selections [8–11,29–32] and chiral states transfer [33–40]. How non-Hermitian periodic systems behave in dynamics thus deserves more attention.

The spin degree of freedom is a critical integrant of the quantum spin Hall (QSH) effect [41–44]. Band connection between time-reversal symmetry (TRS) enforced Kramers pair of spin 1/2 system leads to the $\mathbb{Z}_2$ classification for the QSH effect [41–44]. If TRS is broken by spontaneous magnetization, the quantum anomalous Hall (QAH) effect can appear where only one (pseudo-) spin subspace is topologically nontrivial [45–48]. Non-Hermitian effects such as gain and loss can also break TRS. Their impacts on the QSH effect remain largely unexplored [49,50]. For instance, will there also be QAH-like effects in the presence of (pseudo-) spin-dependent gain and loss, and how will they behave in dynamics? Note here (pseudo-) spin-dependent gain and loss do not necessarily break the TRS of the whole system as TRS relates one spin with gain to another spin with loss, and the system can still preserve TRS.

Here, we consider a breathing honeycomb lattice known to emulate the QSH effect [51–55] and exhibits higher-order corner modes [56,57] for classical waves. We keep the $C_6$ symmetry and introduce non-reciprocal intracell couplings as non-Hermitian terms. Such a system preserves TRS, and the eigenvalues are real or come in complex conjugate pairs. We show that the non-reciprocal intracell couplings exhibit gain for one pseudo-spin subspace and loss with the same magnitude for the other. Thus, in the long-time limit, only one pseudo-spin subspace survives in dynamics, and similar as the QAH, only one helical edge state channel preserves. Effectively, this system exhibits one-way edge states without breaking TRS or using spin-polarized sources. We also find that the spin mixture of the edge modes at the boundaries can be suppressed by the non-reciprocal couplings. In addition, the symmetry protected zero energy higher-order corner modes of a finite system can be completely delocalized.

**Main text**

We start with the BHZ model [41], which can be written as a direct summation of two spin subspaces as $H_{BHZ}(k) = \sum_s \oplus H_s(k)$. Here $H_s(k) = \epsilon_k I + (M + Bk^2)\sigma_z + A(k_x\sigma_x + sk_y\sigma_y)$ and $s = \pm 1$ for two spins. $M$ and $B$ are real numbers, and $A$ is purely imaginary such that TRS is preserved $H_{s=1}(k) = H_{s=-1}^*(-k)$. If $MB < 0$, the BHZ Hamiltonian has a nontrivial topology and exhibits as a QSH system. As shown in Ref. [45], doping with magnetic atoms introduces a spin-dependent term $sG\sigma_z$. When $G > M$, the band is inverted in one of the spin subspaces which leads to a QAH phase. Hence, with the increasing of $G$, one of the spin edge channels merges into the bulk with only one remaining spin edge channel as shown schematically in the upper panel of Fig. 1.

Non-Hermitian terms such as gain and loss are known to break TRS. However, with a proper combination of gain and loss in different spin subspaces, TRS can be preserved. As an example, we consider the non-Hermitian term

$$H_{NH} = isg, \quad (1)$$

which is gain for one spin while loss for the other. Adding $H_{NH}$ to the BHZ model still preserves the time-reversal symmetry, i.e., $H_{s=1}(k) = H_{s=-1}^*(-k)$. However, the eigenvalues now possess an imaginary component $\pm isg$. Hence, the states in one spin channel increase exponentially while those in the other decrease exponentially. In the long-time limit, only states in one spin channel persist, as shown in the lower panel of Fig. 1.

To show the effect of this non-Hermitian term more explicitly, we consider the breathing honeycomb lattice [51–56], as shown in Fig. 2(a). Each primitive cell has six atoms, and we introduce non-reciprocal couplings for clockwise (CW) and counterclockwise (CCW) intracell hopping. The tight-binding Hamiltonian of this non-Hermitian breathing honeycomb lattice reads:

$$H_{NBH} = \sum_{\langle i,j \rangle_{CW}} (t_1 + dt) a_i^\dagger a_j + \sum_{\langle i,j \rangle_{CCW}} (t_1 - dt) a_i^\dagger a_j + \sum_{\langle i,j \rangle} t_2 a_i^\dagger a_j + h.c. \quad (2)$$

where $\langle i,j \rangle_{CW/CCW}$ represent the nearest intracell hopping and $\langle i,j \rangle$ denotes the nearest intercell hopping. When $dt = 0$, such a system is known to possess pseudo-spin Kramers pairs at the $\Gamma$ point, and the effective Hamiltonian near the $\Gamma$ point is the same as the BHZ model [51]. With a proper boundary truncation, that system exhibits helical edge states for both spins. In addition, if that system is finite, it can also possess symmetry-protected higher-order corner modes [56,57]. Before proceeding further, we analyze the symmetry possessed by $H_{NBH}$. First, it has TRS since $H_{NBH} = H_{NBH}^*$. Correspondingly, we have $H_{NBH}(k) = H_{NBH}^*(-k)$, where $H_{NBH}(k)$ is the corresponding Hamiltonian in the momentum space. $H_{NBH}$ also has sublattice symmetry (SLS) $SH_{NBH}S^{-1} = -H_{NBH}$, where $S$ is a $6 \times 6$ matrix with $S_{ij} = \delta_{ij}(-1)^{i-1}$. SLS does not reverse the wave vector. TRS and SLS together lead to a particle-hole symmetry (PHS†) $\mathcal{C}H^*(k)\mathcal{C}^{-1} = -H(-k)$ with $\mathcal{C} = TS$. In addition, $H_{NBH}$ also has a crystalline symmetry $C_6$, thus should exhibit a more complex structure than the 38-fold classification in Ref. [20].

Figure 2(b) shows the real and imaginary parts of the band structure along the high symmetric directions in the Brillouin zone, where the dashed red lines represent the Hermitian case, i.e., $dt = 0$, and the solid blue lines are for $dt = 0.2$. Compared with the Hermitian case, the real parts of the middle four bands for $dt = 0.2$ degenerate in pairs near the $\Gamma$ and K points, and the corresponding imaginary parts are opposite. This feature is protected by the combination of the $C_2$ symmetry and TRS, which requires $C_2 T H(k)(C_2 T)^{-1} = H^*(k)$. This combined symmetry $C_2 T$ requires that the eigenvalues of $H_{NBH}(k)$ are either purely real or come in complex conjugate pairs. This fact can also be seen from the energy spectrum in Fig. 2(c), which is mirror symmetric with respect to the axis $\text{Im}(E) = 0$. In addition, combined with the PHS†, the energy spectrum is also mirror symmetric with respect to $\text{Re}(E) = 0$. If $\Delta t \equiv t_1 - t_2 = 0$, the bands are degenerate at the $\Gamma$ point, and hence the tips of the bow-shape spectra in Fig. 2(c) touch on the $\text{Re}(E) = 0$ line with the specific location determined by $dt$. Otherwise, if $\Delta t \neq 0$, there is always a line gap [19,20] at $\text{Re}(E) = 0$ as denoted by the black dashed line in Fig. 3(c).

Since $C_2T$ also connects the two pseudo-spins, if the eigenvalue of one pseudo-spin of $H(k)$ has a positive imaginary part, the eigenvalue of the other pseudo-spin should exhibit a negative imaginary part with the same magnitude. This point can be seen clearly with the effective Hamiltonian near the $\Gamma$ point when considering only the middle four bands [58–61]

$$H_{eff}(k) = \begin{bmatrix} i\sqrt{3}dt + \Delta t + \frac{1}{4}t_2 k^2 & -\frac{1}{2}ik_- t_2 \\ \frac{1}{2}ik_+ t_2 & i\sqrt{3}dt - \Delta t - \frac{1}{4}t_2 k^2 \end{bmatrix} \oplus \begin{bmatrix} -i\sqrt{3}dt + \Delta t + \frac{1}{4}t_2 k^2 & -\frac{1}{2}ik_+ t_2 \\ \frac{1}{2}ik_- t_2 & -i\sqrt{3}dt - \Delta t - \frac{1}{4}t_2 k^2 \end{bmatrix}. \quad (3)$$

Here $k_\pm = k_x \pm ik_y$, $k^2 = k_x^2 + k_y^2$, and we already block diagonalize the Hamiltonian into the pseudo-spin subspaces [51]. The Hermitian part of $H_{eff}$ is identical to the BHZ model, while the non-Hermitian part exhibits the same form as $H_{NH}$ in Eq. (1). $H_{NH}$ here only shifts the eigenvalues of the pseudo-spin subspaces along the imaginary axis. Thus, the topology of $H_{eff}$ is unchanged with the topological transition point still at $\Delta t = 0$. We show in Supplementary Materials Sec. I. that the projected band structures support this statement.

A nontrivial bulk topology indicates the presence of helical edge modes. However, since the pseudo-spins here are defined under the $C_6$ symmetry which is broken at the boundary, there is unavoidable pseudo-spin flipping, and thus an edge state gaps open. Figure 3(a) shows the semi-infinite lattice with an armchair boundary, where the red and blue arrows denote the propagating direction of pseudo-spins with gain and loss, respectively. Figure 3(b) shows the real parts of the projected band structures at $dt = 0.05$ (left panel) and $dt = 0.2$ (right panel). The color of the edge mode denotes the pseudo-spin component, i.e., whether it is CW or CCW. When $dt$ is small, there is an edge state gap opened at the $\Gamma$ point, as inherited from the Hermitian case. With the increase of $dt$, this gap gradually closed. The edge modes at the $\Gamma$ point actually exhibit a *PT* transition [1], as shown in Fig. 3(c), where we show the real and imaginary parts of the edge state energies at the $\Gamma$ point. The *PT* transition point is at $dt_c = (\sqrt{t_2^2 + 8t_1 t_2} - 2t_1 - t_2)/2$ (See proof in Supplementary Materials Sec. II.) After the transition point, there is no spin mixture at the $\Gamma$ point. Figure 3(d) shows the complex energy spectra for $dt = 0.05$ (left panel), $dt = dt_c$ (middle panel) and $dt = 0.2$ (right panel). Here the edge modes are also colored by their pseudo-spin components. It is clear that CW edge modes always have $\text{Im}(E) > 0$ and CCW edge modes have $\text{Im}(E) < 0$. In addition, edge modes form two loops in the complex energy plane before the transition. These two loops touch at the transition point and eventually merge into a large loop after the transition. In addition, the energy spectra in Fig. 3(d) do not collapse compared to the spectra obtained with the periodic boundary condition in Fig. 2(c), indicating the absence of the non-reciprocal skin effect [6,25].

Previous studies show the presence of symmetry-protected zero-energy corner mode at the 120-degree corner of a finite lattice [41]. Such a corner mode exists inside the gap formed by the two edge modes in Fig. 3(b). When the gap of the edge modes is closed by the non-reciprocal intracell couplings, the corner mode should also be delocalized. We prove in Supplementary Materials Sec. II that $dt_c$ is also the delocalization transition point of the zero-energy higher-order corner state if there is only one corner mode inside a semi-infinite system (See Fig. S2). The situation becomes more complex for a finite system with more than one corner mode. These corner modes interact with each other or with the boundary modes and become delocalized before $dt = dt_c$. Figure 4(a) shows one typical situation where we consider a finite parallelogram shape system with two 120-degree corners and two 60-degree corners. The left panel of Fig. 4(a) shows the real parts of the energy spectra as we increase $dt$. For illustration, we also provide the inverse participation ratio $IPR = \sum_i |\psi_i|^4 / (\sum_i |\psi_i|^2)^2$ in color for the middle four bands. The right panel of Fig. 4(a) shows the amplitude distribution of three typical states [positions marked in the left panel of Fig. 4(a)]. When $dt = 0.05$ (marked by the triangular), the localization length of the corner mode becomes different along the two boundary directions. At a critical value $dt = 0.096$, the corner mode coalesces with another mode, and becomes a fully delocalized edge state thereafter, as shown by the mode at the heart. Such a transition happens before $dt_c$ [the blue vertical dashed line in Fig. 4(a)]. Note that this delocalization process is distinct from the hybrid skin-topological effect [16,62–64] since the latter can either increase or decrease the IPR of the corner states and morph the corner states into an edge state or a bulk state. In contrast, the non-reciprocal coupling in $H_{NBH}$ can only morph the higher-order corner mode into an edge state (see proof in Supplementary materials Sec. II).

When the corner states are delocalized, the helical edge modes extend over the whole complete gap region with one pseudo-spin being amplified and the other decaying in dynamics. Then even if the system is started with a non-spin-polarized source inside the bulk band gap, the edge states with only the amplified spin can survive. However, as shown in Fig. 4(b), the imaginary parts of some bulk states are higher than the edge modes under a uniform $dt$. These bulk states will dominate in dynamics in the long-time limit even with a very small noise, as discussed in *PT* symmetric lasers [8] and non-adiabatic chiral states transfer [33–40]. To address this issue, we only add non-reciprocal coupling for the unit cells on the boundaries, as shown in Fig. 5(a), where $dt = 0.2$ for the red region and $dt = 0$ otherwise. By doing so, the edge states will have the highest Im(E) while the eigenspectra of the edge states and corresponding eigenstates change little compared to the case when non-reciprocal couplings are introduced on the whole area (Detailed discussion in the Supplementary Material Sec. III). When the system is excited with a non-spin-polarized Gaussian source with the center frequency at $\omega_0 = 0$, we can still observe one-way edge mode as shown in Figs. 5(b-f). The dynamics for systems with other boundaries and defects are provided in Supplementary materials Sec. III. Thus, the non-reciprocal coupling can save us from designing dedicate spin polarized source and spin flipping scatterings by purifying the spin in dynamics.

**Summary**

In this work, we investigate the effects introduced by spin-dependent gain and loss using non-reciprocal couplings. Though both spin subspaces are non-Hermitian, the whole system is still time reversal invariant. With the increase of non-reciprocal couplings, the edge mode gap introduced by the spin mixture can be closed, and the zero-energy higher-order corner modes become delocalized. In dynamics, only one spin component can survive in the long-time limit; hence the system behaves as a Chern-like insulator. The tight-binding model we study here can be implemented within various platforms such as circuits [15,65], acoustics [12], and active mechanical systems [17,18], where non-reciprocal couplings are easy to construct. In Supplementary Material Sec. IV, we provide the detailed full-wave simulations of our model using a circuit. Our work provides an intuitive example that the dynamics of a system can be different from the prediction from the static topological invariant. Moreover, our work offers a new direction to manipulate (pseudo-) the spin degree of freedom and morph higher-order corner modes in other condensed matters such as exciton-polaritons [66-69].

**Acknowledgment**

The authors thank Kun Ding for fruitful and helpful discussions. This work is supported by the National Key Research and Development Program of China (Grant No. 2022YFA1404900), the National Natural Science Foundation of China (Grant No. 12274330, 12274332), and Knowledge Innovation Program of Wuhan-Shuguang (Grant No. 2022010801020125).

**Figures**

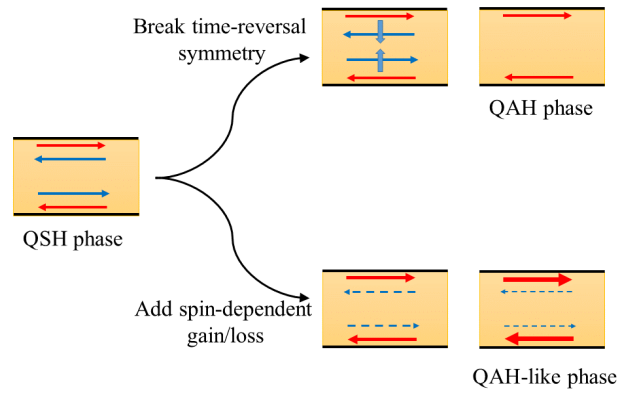

FIG. 1. Sketch shows two approaches to generate spin-polarized one-way edge states starting from a QSH phase. The upper panel shows that under the TRS breaking, the gap of one spin is inverted, and the corresponding edge state channel merges into the bulk, ending in a QAH phase. The lower panel shows that when spin-dependent gain/loss is added, the edge state channel of one spin is amplified, and the other is attenuated, exhibiting a QAH-like phase in the dynamics.

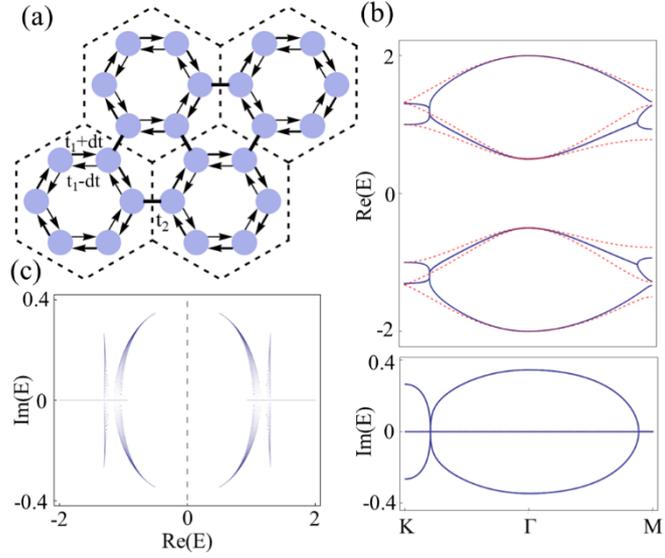

FIG. 2. (a) The tight-binding model. The black dashed lines outline the primitive cells. The intracell couplings are non-reciprocal with $t_1 + dt$ and $t_1 - dt$ for the CW and CCW directions, respectively, and the intercell coupling is $t_2$. (b) The real and imaginary parts of the band structure along high-symmetric directions of the Brillouin zone (BZ). Here the red dashed line, and the blue line correspond to $dt = 0$ and $dt = 0.2$, respectively. (c) Projection of the energy spectrum for states in the BZ. The black dashed line marks the line gap. In (b-c), we keep $t_1 = 0.5$ and $t_2 = 1$.

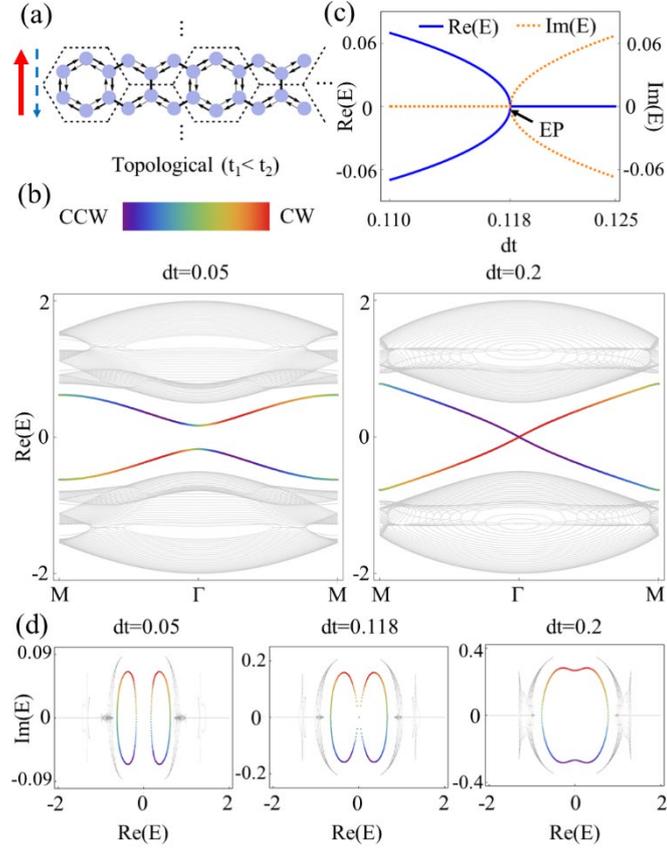

FIG. 3. (a) A supercell used for calculating the edge mode dispersion along the armchair boundary. The solid red (dashed blue) arrow indicates that the CW (CCW) boundary mode is amplified (decaying) in time. (b) The real part of the projected band structures for two different $dt$, with $dt = 0.05$ on the left and $dt = 0.2$ on the right. The gap between the edge modes (colored) is closed at $Re(E) = 0$ with the increasing of $dt$. (c) The edge modes at the $\Gamma$ point exhibit a $PT$ transition with the increasing of $dt$. (d) The energy spectra of the projected band structures at $dt = 0.05$ (left), $dt = dt_c = 0.118$ (middle) and $dt = 0.2$ (right). The edge modes inside the band gap are colored with the chirality index. The CW modes are all above $Im(E) = 0$, while CCW modes are all below $Im(E) = 0$. In (b-d), we set $t_1 = 0.5, t_2 = 1$, and the transition point is at $dt_c = 0.118$.

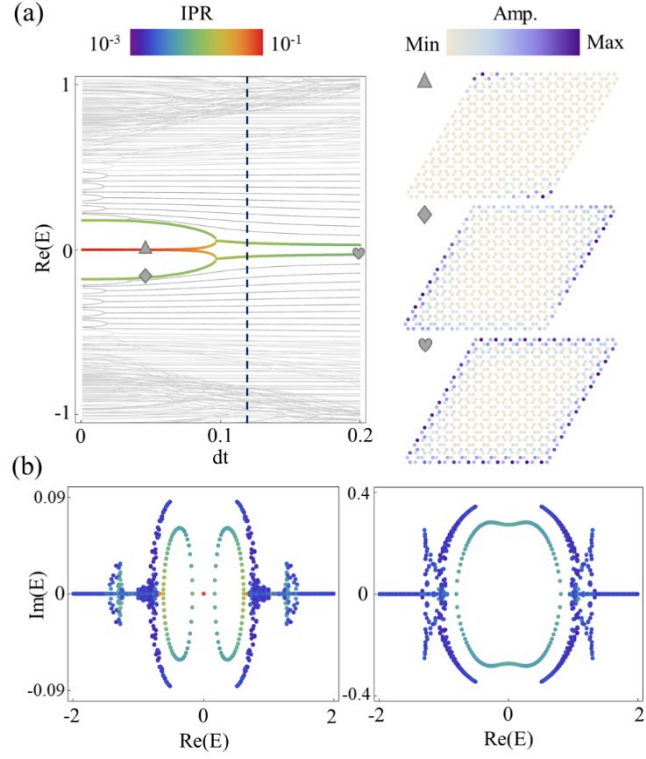

FIG. 4 (a) The real part of the spectra versus $dt$ for a finite parallelogram shape system (shown in the right panel) with four armchair boundaries. IPRs of the middle four modes are given in color. The right panel shows the field amplitude distributions of three representative states (marked in the left panel). The blue dashed line marks $dt = dt_c = 0.118$. (b) The corresponding energy spectra at $dt = 0.05$ (left) and $dt = 0.2$ (right). All the states are colored with their IPRs.

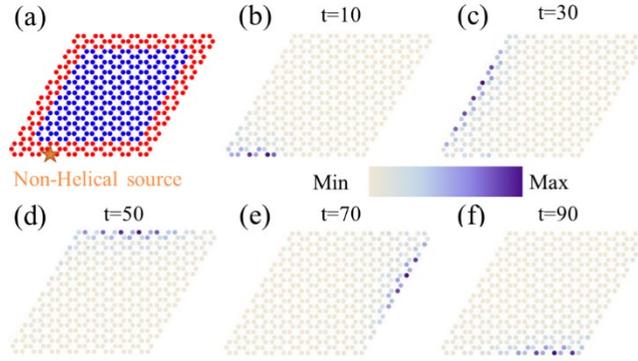

FIG. 5. Dynamics of edge states. (a) The distribution of the non-reciprocal intracell couplings, where $dt = 0.2$ for the red unit cells and $dt = 0$ for the blue ones. The orange star marks the position of the Gaussian point source $S = \exp(i\omega_0 t) * \exp(-(t - t_0)^2/T^2)$ used for exciting the edge states. Here only one lattice site is excited. (b-f) The relative amplitude distributions at different times. The parameters used are $t_1 = 0.5, t_2 = 1, \omega_0 = 0, t_0 = 15, T = 15$. We also add a uniform global loss $\gamma = -0.25$ on every site. For simplicity, we set $\hbar = 1$ in solving the time-dependent Schrödinger equation.